\documentclass[%
reprint,
showpacs,preprintnumbers,
amsmath,amssymb,
aps,
pre,
superscriptaddress
]{revtex4-1}

\usepackage{graphicx}
\usepackage{dcolumn}
\usepackage{bm}
\usepackage{braket}
\usepackage{color}
\usepackage{caption}
\usepackage{ragged2e}
\usepackage{natbib}
\usepackage{hyperref}
\usepackage{subcaption}
\usepackage{url}
\usepackage[utf8]{inputenc}

\DeclareCaptionFormat{myformat}{#1#2(Color online) #3}
\DeclareCaptionJustification{justified}{\justifying} 
{\captionsetup{format=myformat,labelsep=period,justification=justified, font=small}
\figure
}%
{\endfigure}%

\begin{document}
\title{On the functional window of the avian compass}
\author{Vishvendra Singh Poonia}
\affiliation{ Department of Electrical Engineering, Indian Institute of Technology Bombay, Mumbai -- 400076, India}
\author{Kiran Kondabagil}
\affiliation{ Department of Biosciences and Bioengineering, Indian Institute of Technology Bombay, Mumbai -- 400076, India}%
\author{Dipankar Saha}
\affiliation{ Department of Electrical Engineering, Indian Institute of Technology Bombay, Mumbai -- 400076, India}
\author{Swaroop Ganguly}
\email{sganguly@ee.iitb.ac.in}
\affiliation{ Department of Electrical Engineering, Indian Institute of Technology Bombay, Mumbai -- 400076, India}


\date{\today}%

\begin{abstract}
The functional window is an experimentally observed property of the avian compass that refers to its selectivity around the geomagnetic field strength. We show that the radical-pair model, using biologically feasible hyperfine parameters, can qualitatively explain the salient features of the avian compass as observed from behavioral experiments: its functional window, as well as disruption of the compass action by an RF field of specific frequencies. Further, we show that adjustment of the hyperfine parameters can tune the functional window, suggesting a possible mechanism for its observed adaptability to field variation. While these lend strong support to the radical-pair model, we find it impossible to explain quantitatively the observed width of the functional window within this model, or even with simple augmentations thereto. This suggests that a deeper generalization of this model may be called for; we conjecture that environmental coupling may be playing a subtle role here that has not been captured accurately. Lastly, we examine a possible biological purpose to the functional window; assuming evolutionary benefit from radical-pair magnetoreception, we conjecture that the functional window is simply a corollary thereof and brings no additional advantage.
\end{abstract}

\maketitle
\section{Introduction}
\label{Intro}
Avian Magnetoreception -- the ability of some bird species to navigate by sensing (the earth’s) magnetic field -- is one of a set of `quantum biological' phenomena~\citep{engel2007evidence,plenio2008dephasing,levi2015quantum,turin1996spectroscopic,brookes2007could,polli2010conical,nagel2006tunneling,gray2003electron,lambert2013quantum,mohseni2014quantum} where non-trivial quantum effects are thought to play an overt functional role even under warm, dirty conditions~\citep{ritz2000model,kominis2009quantum,cai2010quantum,gauger2011sustained,tiersch2012decoherence,hiscock2016quantum}. Understanding of these phenomena could point the way towards engineering room-temperature quantum biomimetic or quantum information systems.

The radical-pair (RP) model of the avian compass hinges on the dynamics of electron spins on a photo-excited radical pair. These spins can be in the singlet or triplet states (or a superposition thereof) before the radicals recombine. The fraction of recombination product obtained from radical pairs in the singlet state is called the singlet yield which acts as a measure of the geomagnetic field inclination. The RP model has been successful in explaining several of the observed behavioral characteristics~\citep{ritz2000model,schulten1978biomagnetic}. These include: photo-initiated operation~\citep{ritz2000model}, dependence on the inclination and not the polarity of the geomagnetic field~\citep{ritz2000model} and disruption by RF fields~\citep{ritz2004resonance,thalau2005magnetic,ritz2009magnetic,wiltschko2015magnetoreception}. The RP recombination time is thence estimated to be of the order of microseconds~\citep{gauger2011sustained,hiscock2016quantum,bandyopadhyay2012quantum}; the coherence time, which should be larger than this recombination time in order for the geomagnetic field to exercise appreciable effect on the RP spin dynamics, is thus expected to be in the tens of microseconds. It is the long coherence time in a noisy environment that makes this system especially intriguing. The compass action also happens to be extremely sensitive to small RF fields of 1.315 MHz~\citep{thalau2005magnetic,ritz2009magnetic}, which happens to be the Larmor frequency of a free electron. This indicates a spin dynamical mechanism for the avian compass in which one of the electron spins is nearly free. (The other electron spin happens to be subjected to a hyperfine interaction, while both of them interact with the geomagnetic Zeeman field.)

There is, however one characteristic of the avian compass which is not yet completely understood. This is the so-called ‘functional window’~\citep{wiltschko1972magnetic,wiltschko1978further}, which refers to a decrease in the compass sensitivity when the Zeeman field magnitude is outside of a window centred on the geomagnetic field. Behavioral experiments have found that apart from the local magnetic field of 47 $\mu T$, birds are receptive to the magnetic fields of 43 and 54 $\mu T$; however, they get disoriented for 16, 34, 60, 81 and 150 $\mu T$~\citep{ritz2000model}. Moreover, if the bird is exposed long enough to a magnetic field intensity, its compass gets `trained' and is re-centered on the new magnetic field~\citep{ritz2000model}.

An early indication of the dependence of the compass sensitivity, viz. the singlet yield, on the magnetic field was reported by Rodgers and Hore~\citep{rodgers2009magnetic}; later Bandyopadhyay et al.~\citep{bandyopadhyay2012quantum} analyzed the effect on the compass sensitivity of increasing/decreasing the Zeeman field by 30\%~\citep{wiltschko2011mechanism}. Recently, Xu et al. proposed a hyperfine parameter set for the avian compass for which the compass responds to $\pm$30\% change in ambient magnetic (Zeeman) field and external RF field (1.315 MHz) upto the intensity of 15 nT~\citep{xu2013estimating}. From the biological perspective, it seems plausible that the functional window behavior is evolved to enhance compass selectivity to the earth’s magnetic field. 

In this paper, we show that the window-like behavior centered at the local geomagnetic field (as well as the RF disruption) emerges from the RP model for a biologically feasible hyperfine parameter set. We started out with a few hundred such parameter sets, and narrowed down -- first to those which yield a functional window centered on the geomagnetic field of 47 $\mu T$, and then further, to those that result in compass function disruption with a 1.315 MHz RF field. We have also shown that a realistic variation of the compass parameters can explain its adaptive property. Further, we have shown that the compass properties endure for the biologically feasible parameter set even when we relax the usual -- but possibly unrealistic -- condition of equal recombination rates from the singlet and triplet states~\citep{lau2014spin}. The parameter set considered here enables us to predict the range of RP recombination times, and thus, the coherence time of RP spin states -- which turn out to be more than 25 $\mu s$. We show that the functional window behavior is preserved until the environmental noise rate becomes comparable to the recombination rate (k = $4\times10^4s^{-1}$). Finally we explore the possible evolutionary benefit/s from the functional window in the more general context of such benefits from RP avian magnetoreception itself.

The paper has been organized as follows: In Sec.~\ref{RPModel}, we discuss the quantum dynamics of RP model of avian compass including the method of its simulation and explore the functional window in it. In Sec.~\ref{FW}, we explore the functional window characteristic of the avian compass and analyze the trends it follows when various compass parameters are changed. Additionally, we explore the effect of environmental noise on the functional window. In Sec.~\ref{Discussion}, we discuss evolutionary aspects of avian magnetoreception and the functional window, as well as limitations of the RP model. In Sec.~\ref{Conclusion}, we we conclude with an assessment of the RP model as a candidate for explaining the functionality of the avian compass.

\section{The Radical-Pair Model}
\label{RPModel}
The RP model involves a photogenerated radical pair wherein each radical experiences hyperfine interaction with neighboring nuclei. Both radicals interact with the geomagnetic field and therefore the ensuing spin dynamics of the radical pair is influenced by both Zeeman and hyperfine interactions before radicals recombine. The recombination product of the radical pair depends on the spin state of radical pair just before the recombination i.e. singlet and triplet spin states give different/distinguishable chemical products after the radical pair recombine. Also, the biological environment around the radical pair is responsible for dephasing~\citep{walters2014quantum}. The product yield after recombination corresponding to singlet or triplet states contains the information about the magnetic field and both Zeeman and anisotropic hyperfine interactions makes it so~\citep{poonia2015state}. In order to study the functional window and other behavioral characteristics of the avian compass, we choose an illustrative RP system wherein only one of the radicals undergoes hyperfine interaction with a nucleus~\citep{hogben2009possible,cai2012quantum}. Although much more elaborate modeling would be required to simulate the details of the RP mechanism~\citep{hiscock2016quantum}, this model captures the qualitative functionality of the avian compass~\citep{gauger2011sustained}. The RP Hamiltonian is:

\begin{eqnarray}
\label{eq1}
H =\gamma \mathbf{B} \cdot (\hat{S_1} + \hat{S_2}) + \hat{I} \cdot \mathbf{A} \cdot\hat{S_2}
\end{eqnarray}

$\hat{S_1}$ and $\hat{S_2}$ are electron spin operators, and $\hat{I}$ is the nuclear spin operator. $\mathbf{A}$ is the hyperfine tensor and can be written as: $ \mathbf{A}=diag(a_x,a_y,a_z)$. The geomagentic field is characterized by $\mathbf{B}= B_0(sin\theta cos\phi, sin\theta sin\phi, cos\theta)$; $B_0 (= 47 \mu T)$ is the local geomagnetic field at Frankfurt~\citep{ritz2009magnetic} and $\theta$ is the magnetic field orientation with respect to $B_z$ direction which is taken along the RP axis. The photogeneration of the radical pair is the starting point of RP spin dynamics and this is taken to be t = 0. The radical pair is initially in singlet state and nuclear spin state is depolarized~\citep{ritz2000model,timmel1998effects,gauger2011sustained,poonia2015state}. The dynamics of of the RP system is simulated using the master equation approach with quantum toolbox in python (QuTiP) module ~\citep{gauger2011sustained,poonia2015state, johansson2013qutip}. The RP recombination is modeled via Lindblad operators in the master equation (ME) as: $P_1 = \ket{S}\bra{s,\uparrow}$, $P_2 = \ket{S}\bra{s,\downarrow}$, $P_3 = \ket{T_0}\bra{t_0,\uparrow}$, $P_4 = \ket{T_0}\bra{t_0,\downarrow}$, $P_5 = \ket{T_+}\bra{t_+,\uparrow}$ , $P_6 = \ket{T_+}\bra{t_+,\downarrow}$, $P_7 = \ket{T_-}\bra{t_-,\uparrow}$ and $P_8 = \ket{T_-}\bra{t_-,\downarrow}$ where the arrow ($\ket{\uparrow}, \ket{\downarrow}$) are the states of the nucleus. The master equation, then, is given as:
\begin{eqnarray}
\label{eq2}
\dot \rho = - \frac{i}{\hbar}[H, \rho] + k \sum\limits_{i=1}^8 P_i \rho P_i^\dagger - \frac{1}{2}(P_i^\dagger P_i \rho + \rho P_i^\dagger P_i) 
\end{eqnarray}
Where $k$ is the RP recombination rate. The joint state of the radical pair and nucleus at $t=0$ is: $\rho(0) = \frac{1}{2}I \otimes (\ket{s} \otimes \bra{s})$. The singlet yield is defined as the proportion of chemical product after recombination that has singlet precursor. The triplet can also similarly be defined. The variation of the singlet yield $\Phi_S$ (viz. the fraction of radical pairs recombining from the singlet spin state) with the geomagnetic field inclination leads to compass functionality in the RP model. Most of the following discussion will, in fact, be in terms of the compass sensitivity $D_S=\Phi_S^{max} - \Phi_S^{min}$, viz. the difference between the maximum and minimum singlet yield as a function of inclination~\citep{cai2012quantum}.

\begin{figure}
\centering
\includegraphics[width=.8\linewidth]{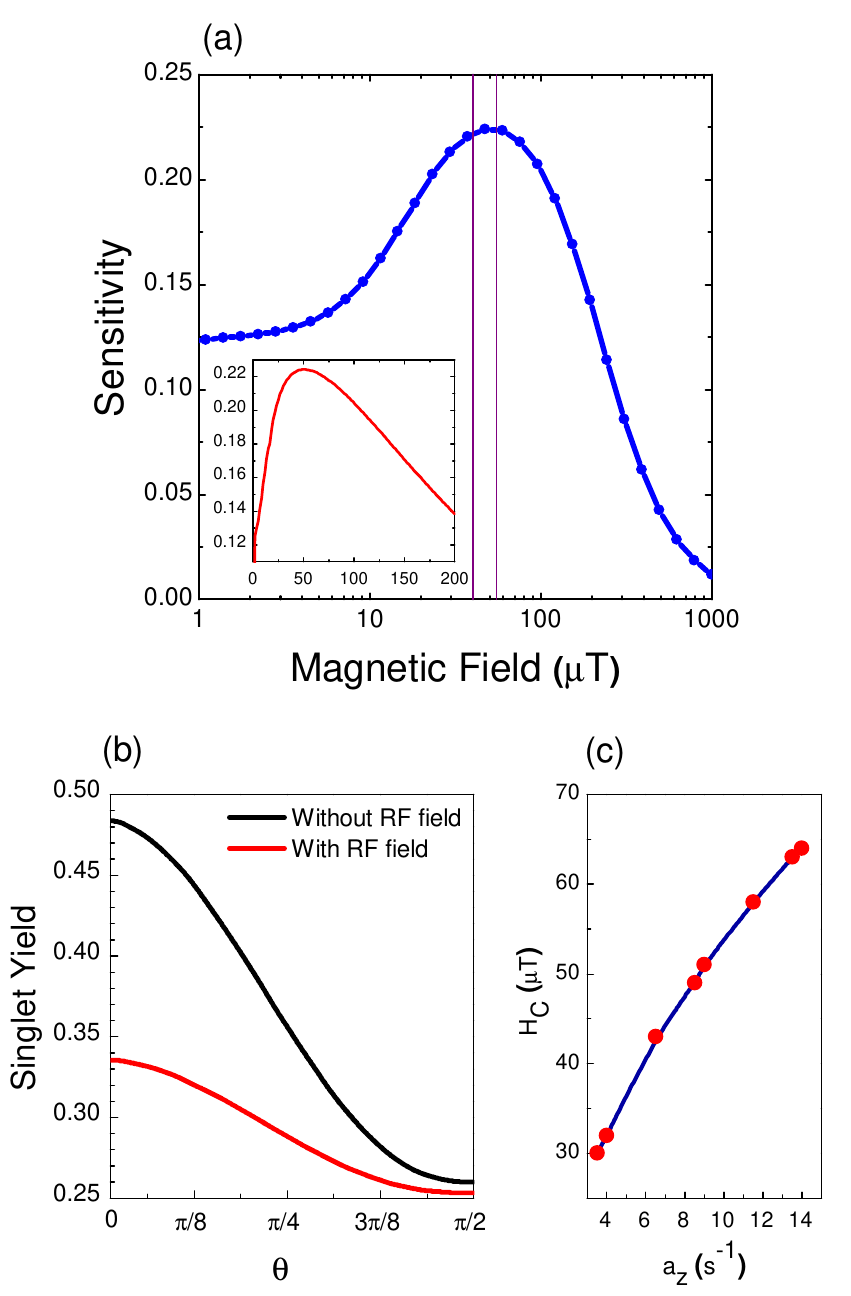}
\caption{(color online) (a) Compass sensitivity as a function of Zeeman field for ($a_x,a_y,a_z,k_s,k_t$)=(0.345 G, 0.345 G, 9 G, $2 \times 10^4 s^{-1}, 2 \times 10^4 s^{-1}$). The plot clearly reveals a `functional window' centered around 47 $\mu T$.  Behavioral experiments show compass sensitivity only between the vertical (purple) lines, i.e. a far narrower functional window. The inset displays the functional window for linear scale of magnetic field. (b) Singlet yield as a function of geomagnetic field inclination for ($a_x,a_y,a_z,k_s,k_t$)=(0.345 G, 0.345 G, 9 G, $2 \times 10^4 s^{-1}, 2 \times 10^4 s^{-1}$) with [red] and without [black] 150 $nT$ RF field of 1.315 MHz frequency. The figure distinctly shows the degradation of compass sensitivity in presence of RF field, a defining characteristic of the avian compass. (c) Variation of center of the functional window (denoted as $H_C$) as a function of axial hyperfine strength ($a_z$) -- indicative of adaptive behavior.}
\label{Fig1_FW}
\end{figure}

\section{The functional window}
\label{FW}
The functional window is defined as band pass filter like characteristic wherein the compass sensitivity is appreciable only for a narrow range of magnetic fields and negligible otherwise. In order to locate the functional window, the compass sensitivity as a function of the geomagnetic (Zeeman) field intensity is analyzed for a large number of hyperfine parameters and RP recombination rates. The sharpest functional window centered at 47 $\mu T$ then obtains for hyperfine parameter set of $(a_x, a_y, a_z) = $(0.345 G, 0.345 G, 9 G) with $(k_s, k_t) = $($2\times10^4s^{-1}, 2\times10^4s^{-1})$; this is shown in Fig.~\ref{Fig1_FW} (a). This parameter set is seen to fall within the biologically-feasible regime of hyperfine interaction strength, which ranges from 0.1 G - 10 G~\citep{rodgers2009magnetic}. Moreover, this set of hyperfine and recombination rate parameters also exhibit the RF disruption property, as shown in Fig.~\ref{Fig1_FW} (b). The RF field of 1.315 MHz is considered to be disrupting the avian compass functionality if the compass sensitivity drops by more than 30\% in presence of the RF field~\citep{bandyopadhyay2012quantum,wiltschko2011mechanism}. The singlet yield with and without RF field is shown in Fig.~\ref{Fig1_FW} (b) for the aforementioned hyperfine and recombination parameter set which clearly illustrates the RF disruption of avian compass.

\begin{figure}
\centering
\includegraphics[width=0.80\linewidth]{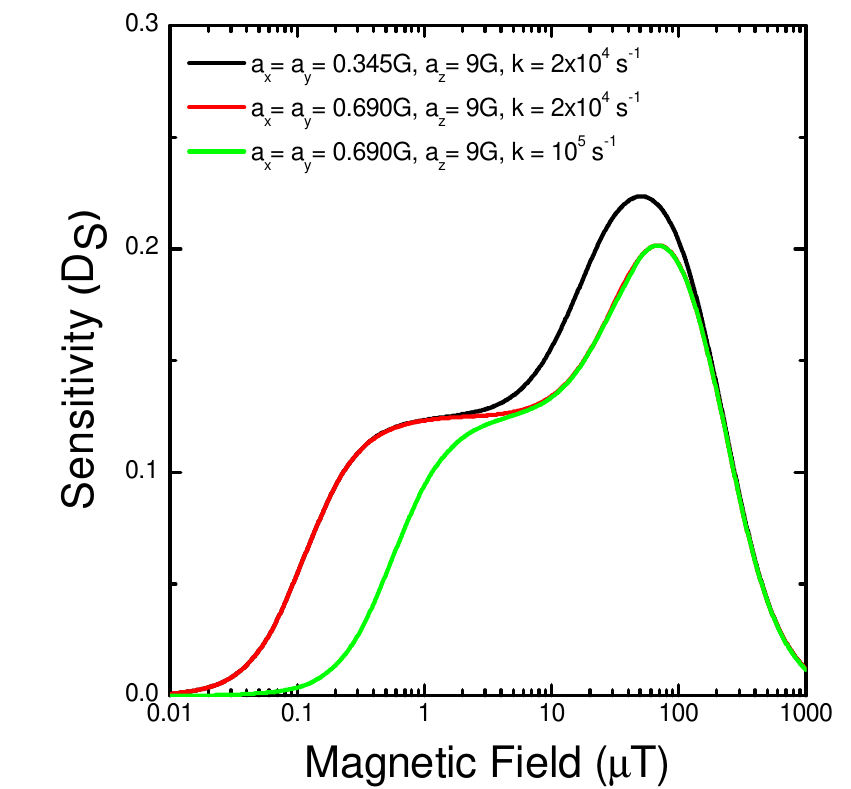}
\caption{(color online) The general trend of functional window with hyperfine and recombination rate parameters is shown here. Functional window is plotted for the parameter values of ($a_x,a_y,a_z,k_s,k_t$)=(0.345 G, 0.345 G, 9 G, $2 \times 10^4 s^{-1}, 2 \times 10^4 s^{-1}$), ($a_x,a_y,a_z,k_s,k_t$)=(0.690 G, 0.690 G, 9 G, $2 \times 10^4 s^{-1}, 2 \times 10^4 s^{-1}$), ($a_x,a_y,a_z,k_s,k_t$)=(0.690 G, 0.690 G, 9 G, $ 10^5 s^{-1}, 10^5 s^{-1}$). $a_z$ is kept constant in these plots. The plots show that on increasing (decreasing) the value of $a_x$ and $a_y$, the height of functional window decreases (increases). An increment in the recombination rate shifts the lower portion of the functional window to higher values of magnetic field.}
\label{Fig2_FWTrends}
\end{figure}

Henceforth, we discuss the the methodology of discovering the functional window and the regime of hyperfine and recombination parameters explored . The z-axis is assumed along the RP axis. The hyperfine parameter sets can broadly be divided into two regimes: cigar-shaped  ($a_x = a_y < a_z$) and disk-shaped ($a_x = a_y > a_z$), assuming symmetry in the transverse plane. For cigar-shaped hyperfine parameters, we examine $a_z$ value varying from 0 to 100$B_{geo}$ and $a_x$ and $a_y$ are varied from 0 to $a_z$. Similarly for disk-shaped hyperfine parameters, the values of $a_x$ and $a_y$ are varied from 0 to 100$B_{geo}$ and $a_z$ is varied from 0 to $a_x (= a_y)$. In addition to this, we also explored these two set of hyperfine parameters for the case when $a_x \neq a_y$ but it didn't offer any distinctive observation. For these hyperfine parameters, recombination rates ($k_s,k_t$) from $10^4 s^{-1}$ to $10^7 s^{-1}$ were examined. Functional window having varied width and center at different Zeeman magnetic field is obtained for many combination of these parameters and a general trend is observed with respect to the strength of hyperfine parameters and recombination rates. Qualitatively following things are observed: a) As the value of $a_z$ increases, the center and width of the functional window shifts towards higher values of Zeeman field. The lower part of the functional window remains still with respect to the change in the value of $a_z$. b) $a_x$ and $a_y$ only affect the height of the functional window and not the width. Higher the values of $a_x$ (or $a_y$), lower the height of the functional window. c) The recombination rate affects the lower limit of the functional window and increasing (decreasing) the recombination rate would shift the lower limit of the functional window towards higher (lower) magnetic field strengths. However, the constraint associated with the recombination rate is that the compass starts loosing its RF disruption property as the recombination rate is increased. The general trends of functional window with respect to hyperfine interaction strength and recombination rates are captured in Fig.~\ref{Fig2_FWTrends}. During the exploration of functional window, the parameter set was narrowed down by setting the requirement of a peak around the geomagnetic field with functional window as narrow as possible and display of the the RF disruption property. The parameter set: $(a_x, a_y, a_z, k_s, k_t) = $(0.345 G, 0.345 G, 9 G, $2\times10^4s^{-1}, 2\times10^4s^{-1}$) exhibit both functional window and the RF disruption property. 

The analysis suggests that the functional window is observed around the local geomagnetic field if $a_x$ and $a_y$ are small in comparison to $a_z$ (cigar-shaped hyperfine interaction with $a_x = a_y << a_z$). However if they happen to be vanishingly small ($a_x = a_y \approx 0$), the functional window becomes relatively much broader in the lower magnetic field regime, something that is not observed in behavioral experiments. For larger values such that $a_x = a_y > a_z$, we still get a functional window but the sensitivity (functional window height) is very small. For still larger values such that $a_x \approx a_y \approx a_z$, the functional window is completely washed out. In contrast, the functional window turns out to be largely insensitive to the recombination rates -- to the extent that increasing or decreasing these $(k_s, k_t)$ upto a factor of 8 with respect to the optimal value does not change the window appreciably. However, this value of recombination rate parameters models the RF disruption properly as demonstrated in Fig.~\ref{Fig1_FW} (b).

From Fig.~\ref{Fig1_FW} (a), we also observe that while the RP model clearly captures the functional window behavior qualitatively, it is unable -- even in this best case -- to reproduce its experimentally observed width and sharpness quantitatively. We shall revisit this point in a bit.

\begin{figure}[t]
\centering
\includegraphics[width=1.00\linewidth]{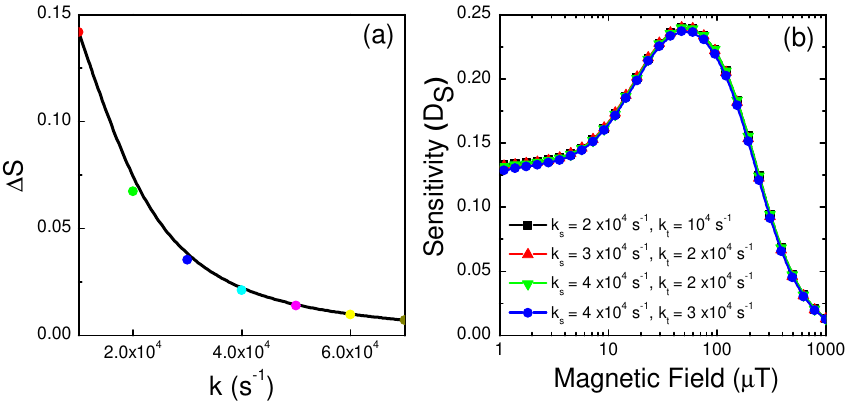}
\caption{(color online) (a) Change in sensitivity of the compass as a function of RP recombination rate. (b) Sensitivity as a function of Zeeman field strength when singlet and triplet recombination rates are different. We take $k_s > k_t$ which is implied by RP theory~\citep{lau2014spin}. The curve is plotted for ($k_s, k_t$) = ($2 \times 10^4 s^{-1}$, $10^4 s^{-1}$),($3 \times 10^4 s^{-1}$, $2 \times 10^4 s^{-1}$), ($4 \times 10^4 s^{-1}$, $2 \times 10^4 s^{-1}$),($4 \times 10^4 s^{-1}$, $3 \times 10^4 s^{-1}$).}
\label{Fig3_FW}
\end{figure}

We now point out that not only does the essential functional window behavior emerge from the RP model, but so does its adaptability characteristic. Fig.~\ref{Fig1_FW} (c) shows the variation of center of functional window for various hyperfine interaction strengths. As the magnitude of the hyperfine interaction ($a_z$) increases (decreases), the center of functional window shifts towards higher (lower) magnetic field values. We note that it may be this capability that lets avian species adjust to spatio-temporal variations of the geomagnetic field~\citep{mcelhinny1998magnetic}. Physically, the hyperfine interaction strength has inverse relation with the distance ($1/r^3$) between nucleus and electron. Therefore, the adaptive behavior here could be realized by small structural adjustments in the radical pair that could modify the hyperfine parameters. Further this behavior, like the functional window itself, is again largely insensitive to changes in the recombination rates $(k_s, k_t)$. 

As stated above, the RF disruption property was analyzed for various values of recombination rates. Fig.~\ref{Fig3_FW} (a) shows the change in sensitivity as a function of recombination rate, $k$ $(k_s,k_t)$. It is found that only $k_s, k_t$ values less than $4\times10^4s^{-1}$ lead to RF disruption in this sense, and are therefore acceptable from the RP model perspective. From these values of the recombination rate, we conclude that the coherence time of the radical pair must be at least $(4 \times 10^4)^{-1}s = 25 \mu s$. This rather large time-scale seems to corroborate the quantum nature of the avian compass and suggests that it may provide useful learning for quantum technologies.

Next, we analyze the functional window when the recombination rates for the singlet and triplet channels, $k_s$ and $k_t$, are different. It is usual for these to be considered identical~\citep{gauger2011sustained, bandyopadhyay2012quantum}, but that is actually not to be expected from the physical basis of RP theory. The radical pair can readily recombine back if it happens to be in the singlet state but not from the triplet, whence it can only form escape products with other species~\citep{lau2014spin}. This suggests that the recombination rate of singlet radical pairs should be taken more than the recombination (or, more accurately the escape) rate of triplet radical pairs i.e. $k_s > k_t$. Fig.~\ref{Fig3_FW} (b) shows the `functional window' behavior of the compass when $k_s > k_t$, with both parameters coming from the range $(k_s, k_t \leq 4\times10^4s^{-1})$. We observe that the functional window is practically unchanged even when the singlet and triplet recombination times are treated realistically thus. We point out that the RF disruption property is also preserved as long as $k_s$ and $k_t$ are within the aforementioned range, albeit unequal.

\begin{figure}
\centering
\includegraphics[width=.8\linewidth]{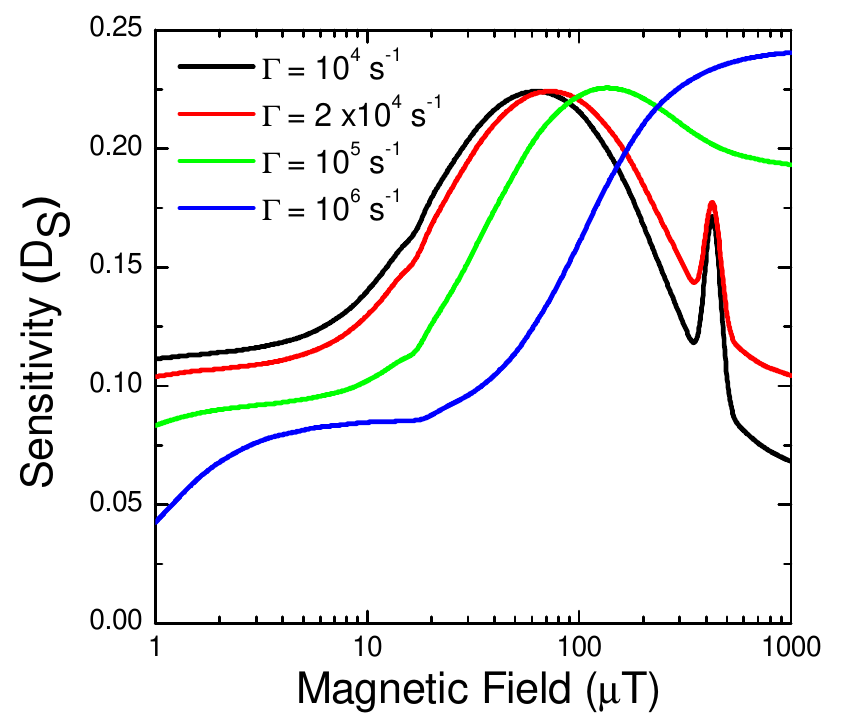}
\caption{(color online) Compass Sensitivity as a function of Zeeman field strength in presence of environmental noise~\citep{walters2014quantum}. The sensitivity is analyzed for $(a_x, a_y, a_z, k_s, k_t) = $(0.345 G, 0.345 G, 9 G, $2\times10^4 s^{-1}, 2\times10^4 s^{-1})$ along with environmental noise rates ($\Gamma$) of $10^4 s^{-1}$, $2\times10^4 s^{-1}$, $10^5 s^{-1}$, $10^6 s^{-1}$. As is clear from the figure, the functional window property vanishes when the noise rate is greater than the RP recombination rate.}
\label{Fig4_FWNoise}
\end{figure}

Further, we explore the effect of environmental noise on the functional window using a projection noise model~\citep{walters2014quantum}. Fig.~\ref{Fig4_FWNoise} shows functional window for $(a_x, a_y, a_z, k_s, k_t) = $(0.345 G, 0.345 G, 9 G, $2\times10^4 s^{-1}, 2\times10^4 s^{-1})$ in presence of noise rates of $10^4 s^{-1}, 2\times10^4 s^{-1}, 10^5 s^{-1}, 10^6 s^{-1}$. Expectedly, it shows that the functional window vanishes for noise rates that are larger than the recombination rate/s. We note the presence of small spikes in the sensitivity beyond the main peak for small values of the noise rate. Whether these have any special significance~\citep{poonia2015state} is not clear to us at this point and the .

\section{Discussion}
\label{Discussion}
What we have presented so far argues strongly in favor of the RP model. However, the following fundamental question is still left open: what purpose, if any, does the functional window of the avian compass actually serve? We begin by reviewing what we know about the evolutionary purpose behind avian magnetoreception itself, in particular radical-pair based magnetoreception. It is accepted that cryptochrome molecules are the most probable source of radicals in the RP mechanism~\citep{liedvogel2007chemical,ritz2000model} Cryptochromes have been  discovered in plants, birds, animals and worms ~\citep{liedvogel2007chemical, lin2005cryptochromes, occhipinti2014magnetoreception}; they are essential for growth and development in plants, circadian clock and magnetoreception in animals. It is hypothesized that the animal and plant cryptochromes have evolved from certain kind of photolyases, probably through gene/genome duplication and evolved different functions~\citep{lucas2009evolution}. These photolyases repair UV-induced DNA damage by a photo-induced cyclic electron transfer mechanism. It is further suggested that at least four different forms of photolyases were present in the common ancestor of all three forms of cellular life - bacteria, archaea and eukarya; and plant and animal cryptochromes evolved different functions~\citep{lucas2009evolution}. In addition, many studies on the evolutionary origins of photolyases showed that photolyases that lead to the emergence of cryptochromes were present in prokaryotes before the emergence of eukaryotes~\citep{lucas2009evolution}. This suggests that all forms of life, right from the beginning had the evolutionary potential for magnetoreception. However, the role of geomagnetic field as an influential abiotic evolutionary factor is  largely unexplored and may be debatable. It could conceivably be addressed by considering magnetoreception as a selection force conferring some specific advantage, and finding common timelines for the evolution of the earth's magnetic field over the last four billion years (perhaps by looking at iron-deposition data) and the evolution of cryptochromes; one would also need to consider migration of animals, both local and long distance, that was necessitated by the burgeoning population and seasons, especially after the Cambrian explosion about 550 million years ago that led to rapid speciation and population explosion. Here we make a provisional assumption that magnetoreception itself might be providing some evolutionary benefit by aiding migration, and thereafter examine whether a functional window therein brings any further advantage. We first consider an apparently plausible evolutionary purpose for the large-field (right-hand) side of the window, viz. to protect this magnetic field sensor from large stray magnetic fields and fluctuations. However, major fluctuations like solar flares that do disorient certain bird species~\citep{WinNT1,WinNT2} have very small amplitudes -- it is likely that the disruption in fact occurs via its RF disruption property~\citep{ritz2004resonance}. On the other hand, spatial (pole-equator) and slow temporal (secular, i.e. roughly annual) variations of the geomagnetic field -- that birds obviously need to be able to sense -- are actually larger than the experimentally observed functional window. All things considered, it then seems plausible that the functional window is an incidental feature, a by-product of the RP spin dynamics, with no obvious evolutionary benefit (we note that the question of evolutionary benefit has come up for debate in other areas of quantum biology as well~\citep{ball2004enzymes,nagel2006tunneling}). In such a case, realizing an overly wide window of operation -- to accommodate spatio-temporal field variation -- would add `cost' (complexity) that seems to have been eschewed by biology; it has, in fact, gone for a sharper window than we predict from the RP model, choosing to manage field variation by having the window adaptive, which, as we have shown here, is also a natural consequence of the RP dynamics. Since a narrower functional window does not even serve any apparent purpose, it seems reasonable to rule out resource investment in additional `sense-amplification circuitry' to achieve window width sharper what we get `for free' from the RP dynamics. We are thus left with no explanation for the sharper-than-predicted functional window within the RP model -- and should therefore look without. We posit the possibility that this model -- for its considerable successes including those shown here -- may not be complete yet; when it is, the experimentally observed sharp functional window should emerge from the RP spin dynamics itself. We note here that we attempted to generalize the RP model by incorporating spin-spin interactions, namely exchange and dipolar, but found that these do not help to model the sharpness of the functional window; this agrees with earlier model predictions that the RP spin dynamics is robust against such external interactions~\citep{efimova2008role,dellis2012quantum}. That in turn means that hybrid mechanisms, combining RP and magnetic particles~\citep{qin2016magnetic,lohmann2016protein}, would also not materially affect the functional window. We feel that the necessary generalization of the RP model may be in terms of subtle environmental interaction, which has a well-known precedent in quantum biology~\citep{plenio2008dephasing,chin2010noise}.

\section{Conclusion}
\label{Conclusion}
In conclusion, we have shown that the qualitative characteristics of the avian compass emerge from the RP model with biologically feasible parameters. On the face of the aggregate evidence, it seems fair to say that the RP model is nearly, but not yet definitively, established as the mechanism for the avian compass. In particular, we have shown that it leads to the qualitative functional window behavior for a highly anisotropic, hyperfine parameter set, but in its present form, cannot quantitatively match the experimentally observed sharpness of the window. We conjecture that this may be a shortcoming of the way that environmental interactions have been modelled so far. We further observe that the behavioral property of adaptability (to a different Zeeman field) can also be easily explained within the RP model through moderate tuning of the hyperfine parameters. The same hyperfine parameters together with appropriate recombination parameters also lead to the RF disruption property of the avian compass -- which is found to be more sensitive to the latter. The functional window, on the other hand, is found to be generally insensitive to variation of recombination rates and to unequal singlet and triplet recombination rates. The recombination rates, however, are essential in setting the coherence time for the system which in turn decides the level of environmental noise it can withstand. We show that for noise rates larger than the recombination rate, the functional window vanishes. Finally, we observe that even if we assume evolutionary benefits accruing to bird species from RP magnetoreception, there is no clear biological raison d'etre to have a the functional window therein; thus, it could simply be a by-product of the RP spin dynamics, with no utilitarian role per se. \\ \\

\acknowledgments{We are grateful for the support from the Department of Electronics and Information Technology through the Centre of Excellence in Nanoelectronics at IIT Bombay.}

\bibliography{FunctionalWindow}
\end{document}